\documentstyle[prl,aps,epsf,multicol]{revtex}

\begin{document}
\draft
\tighten

\title{``Soft'' Anharmonic Vortex Glass in Ferromagnetic Superconductors}
\author{Leo Radzihovsky$^1$, A. M. Ettouhami$^1$, 
Karl Saunders$^2$, John Toner$^2$}

\address{$^1$ Department of Physics, University of Colorado,
Boulder, CO 80309}
\address{$^2$ Dept. of Physics,
Materials Science Inst., and Inst. of Theoretical Science, University
of Oregon, Eugene, OR 97403}

\date{\today}
\maketitle
\begin{abstract}
Ferromagnetic order in superconductors can induce a {\em spontaneous}
vortex (SV) state.  For external field ${\bf H}=0$, rotational
symmetry guarantees a vanishing tilt modulus of the SV solid, leading
to drastically different behavior than that of a conventional,
external-field-induced vortex solid.  We show that quenched disorder
and anharmonic effects lead to elastic moduli that are
wavevector-dependent out to arbitrarily long length scales, and
non-Hookean elasticity. The latter implies that for weak external
fields $H$, the magnetic induction scales {\em universally} like
$B(H)\sim B(0)+ c H^{\alpha}$, with $\alpha\approx 0.72$.  For weak
disorder, we predict the SV solid is a topologically ordered vortex
glass, in the ``columnar elastic glass'' universality class.
\end{abstract}
\pacs{64.60Fr,05.40,82.65Dp}

\begin{multicols}{2}
\narrowtext

Rare-earth borocarbide materials exhibit a rich phase diagram that
includes superconductivity, antiferromagnetism, ferromagnetism and
spiral magnetic order.\cite{PhysicsToday,experiments,Varma}
In particular, there is now ample experimental evidence that, at low
temperatures, superconductivity and ferromagnetism competitively
coexist in ErNi$_2$B$_2$C compounds. Other possible examples of such
ferromagnetic superconductors (FS) are the recently discovered high
temperature superconductor Sr$_2$Y Ru$_{1-x}$Cu$_x$O$_6$ and the
putative $p$-wave triplet strontium ruthenate superconductor, Sr$_2$Ru
O$_4$, which spontaneously breaks time reversal symmetry.
For sufficiently strong ferromagnetism, such FS's have been
predicted\cite{Varma} to exhibit a {\em spontaneous} vortex (SV) state
driven by the spontaneous magnetization, rather than by an external
magnetic field ${\bf H}$.  The novel phenomenology of the associated
SV solid is the subject of this Letter.

\begin{figure}[bth] 
\centering
\setlength{\unitlength}{1mm} 
\begin{picture}(25,60)(0,-12)
\put(-60,-50){\begin{picture}(20,20)(0,0) 
\includegraphics{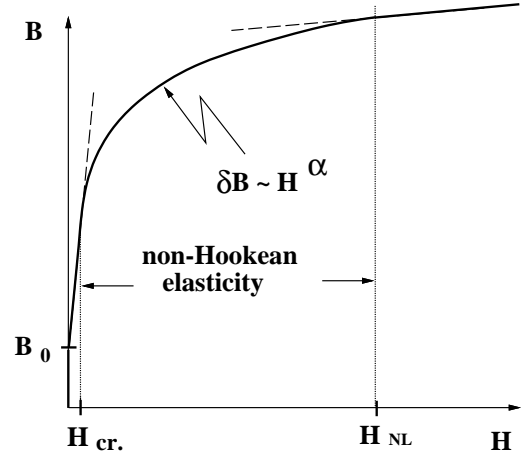} 
\end{picture}} 
\end{picture} 
\caption{The nonlinear and universal power-law $B(H)$ scaling, 
which at weakest fields $H<H_{cr.}$ and strongest fields $H>H_{NL}$
is cutoff by the crystal symmetry breaking anisotropy and $\xi_{NL}$,
respectively.}
\label{BvsH} 
\end{figure}

Here we will show that for ${\bf H}=0$, the elastic properties of the
resulting SV solid differ dramatically and {\em qualitatively} from
those of a conventional Abrikosov lattice. The key underlying
difference is the {\em vanishing} of the tilt modulus, which is
guaranteed by the underlying rotational invariance (but see
below). Although this invariance is broken by the magnetization, ${\bf
M}$, the tilt modulus remains zero because this breaking is {\em
spontaneous}.  This contrasts strongly with a conventional vortex
solid, where the rotational symmetry is {\em explicitly} broken by the
{\em applied} field ${\bf H}$.  All of our conclusions, e.g., the
unusual $B(H)$ relation, illustrated in Fig.\ref{BvsH} are a direct
consequence of this important observation.


In particular, we find that this ``softness'' (i.e., vanishing tilt
modulus) of the SV lattice {\em drastically} enhances the effects of
quenched disorder. As in conventional vortex lattices\cite{Larkin},
{\em any} amount of disorder $\Delta_V$, however weak, is sufficient
to destroy translational order in SV lattices. Here the finite ordered
domains are {\em divergently anisotropic}, with dimensions
$\xi^L_\perp\propto 1/\Delta_V^{2/3}$ and $\xi^L_z\propto
1/\Delta_V^{1/3}$.  These lengths are measurable in scattering and
transport measurements.  Unlike conventional lattices, however, in SV
lattices the disorder also qualitatively alters the elastic behavior
at long distances, leading to ``anomalous elasticity'': a {\em
universal} scaling of elastic moduli with wavevector $\bf k$ out to
{\em arbitrarily} long length scales, with some elastic moduli
vanishing, and others diverging, as wavevector ${\bf k}\rightarrow
0$.\cite{Brandt}
This behavior is characteristic of a new kind of topologically ordered
``columnar elastic glass''(CEG) phase of vortices, which is stable, for
weak disorder, against proliferation of dislocations.
%

The best way to experimentally probe a {\em spontaneously} broken 
symmetry is to break it directly with an external field. 
We predict that as a consequence of the anomalous elasticity, the increase
in the magnetic induction,
$\delta B(H)
\equiv B(H)-B(0)$ 
over the {\em spontaneous} induction $B(0)$, due to a weak
applied field ${\bf H}$ along ${\bf M}$
obeys a {\em universal} ``non-Hookean'' scaling law:
\begin{equation}
\delta B(H) \propto |H|^{\alpha}\;,
\label{nonhook}
\end{equation}
with the {\em universal} exponent $\alpha=0.72 \pm
0.04$,
a prediction that should be experimentally testable.
%
%

Whatever the microscopic origin of the novel FS state, general symmetry
principles dictate that the long length scale phenomenology is described
by an effective Landau-Ginzburg free energy functional
\begin{eqnarray}
\hspace{0cm}F_{GL}&=&\int \! d^3 r\bigg[{\hbar^2\over 2m}|(\bbox{\nabla}- i
{2\pi\over\phi_0} {\bf A})\psi|^2+{a\over2}|\psi|^2+
{b\over4}|\psi|^4\nonumber\\
&+&{K\over2}|\bbox{\nabla}{\bf M}|^2+{t\over2}|{\bf M}|^2+
{u\over4}|{\bf M}|^4+{1\over8\pi}|{\bf B}-4\pi{\bf M}|^2\nonumber\\
&-&{1\over4\pi}{\bf B}\cdot{\bf H}\bigg]\;,
\label{F_GL}
\end{eqnarray}
where $\psi$ is the superconducting order parameter and ${\bf A}$ is the
electromagnetic vector potential with $\bf B=\bbox{\nabla}\times{\bf
  A}$.
The constants $m,a,b,K,t$, and $u$ are experimentally measurable
phenomenological parameters, with $a$ and $t$ changing their signs at
the superconducting and ferromagnetic decoupled transition temperatures
respectively, and $\phi_0=hc/2e$ the elementary magnetic flux quantum.

The above model displays rich phenomenology\cite{Varma}. It has been
shown in detail by Greenside, et al.\cite{Varma}, that for a range of
physically realistic parameters (e.g., $\lambda_L/\xi$,
$\lambda_L/\sqrt{K/|t|}\approx O(10)$,\cite{Varma} with the large
Abrikosov ratio required for a robust mixed state, and large exchange
$K$ necessary to suppress the competing spiral phase, relative to the
SV state), systems described by $F_{GL}$, Eq.\ref{F_GL}, exhibit a
thermodynamically stable phase consisting of a {\em spontaneous}
(i.e. ${\bf H}=0$) vortex state, in which superconducting vortices are
generated by the spontaneous magnetization of the ferromagnetic
component.

Deep within this state, both $\psi({\bf r})$ and ${\bf M}({\bf r})$
are large, and therefore the London approximation applies. We take
${\bf M}({\bf r})=M_0\hat{\bf n}({\bf r})$ to be of approximately
uniform magnitude $M_0$
and fluctuating direction $\hat{\bf n}({\bf
r})$. We also take $|\psi({\bf r})|$ to be a constant $\psi_0\neq0$,
everywhere except at the locations of the vortices (where $\psi=0$)
and allow its phase $\theta({\bf r})$ to vary subject to the
circulation condition that it increases by an amount $2\pi$ along a
closed path enclosing a vortex\cite{ReviewsSC}.  The magnetic
induction ${\bf B}({\bf r})$ is confined to thin flux filaments
defined by the zeros of $\psi$. Using the circulation condition and the
minimizing above free energy
$F_{GL}$ with respect to $\psi_0$, $\theta$ and ${\bf A}$, we
eliminate $\bbox{\nabla} \theta$ and ${\bf B}$ in favor of vortex line
conformational degrees of freedom and obtain:

%
\begin{eqnarray}
F_L&=&{\phi_0^2\over8\pi}\int d^3r d^3r'{\bf t({\bf r})}\cdot{\bf
t({\bf r}')}V(|{\bf r}-{\bf r}'|)\nonumber\\
&+&\int d^3r\left[{1\over2}K M_0^2|\bbox{\nabla}\hat{\bf n}|^2-
\phi_0 M_0{\bf t}({\bf r})\cdot\hat{\bf n}\right]\;,
\label{F_L}
\end{eqnarray}
where we have dropped unimportant constant pieces. ${\bf t({\bf
r})}=\sum_n\int ds {d{\bf r}_n\over ds}\delta^3({\bf r}-{\bf r}_n(s))$
is the directed vortex line density, with magnitude given by the local
flux density (in units of $\phi_0$) and with direction tangent to the
local vorticity; ${\bf r}_n(s)$ is the locus of the $n$-th vortex
line, parameterized by an arc length $s$. The vortex interaction
$V(r)$ is approximately given by the standard expression
$V(r)=e^{-r/\lambda_L}/(4\pi\lambda_L^2 r)$, with $\lambda_L^2=
m\phi_0^2/(16\pi^3\hbar^2|\psi_0|^2)$\cite{Varma}, cut off at scales
smaller than the coherence length $\xi$.

At low temperature $T$ and in the clean limit the vortex
lines freeze into a lattice (as has been observed\cite{PhysicsToday}),
directed on average along a {\em spontaneously} chosen magnetization 
direction $\hat{\bf n}_0$, which we choose to be the $\hat{\bf z}$ axis, 
and with the lattice constant determined by the average magnetization 
$M_0$. Vortex lattice fluctuations are described by the 2d phonon
displacement vector field ${\bf u}(z,{\bf R}_n)$, defined by ${\bf
r}_n(z)=(X_n+u_{x}(z,{\bf R}_n),Y_n+u_{y}(z,{\bf R}_n),z)$, where
${\bf R}_n=(X_n,Y_n)$ spans the vortex lattice, and we chose $s=z$,
and by a small 2d vector $\delta{\bf n}\perp \hat{\bf z}$
defined by $\hat{\bf n}=(1-\delta{\bf n}\cdot\delta{\bf
n})^{1/2}\hat{\bf z}+\delta{\bf n}$.

Integrating out $\delta{\bf n}$ and going to the continuum, we
obtain\cite{Lubensky,unpublished} the effective elastic Hamiltonian:
\begin{equation}
H_{el}={1\over2}\int d^3r\left[\kappa|\partial^2_z{\bf u}|^2+2\mu
u_{ij}^2 +\lambda u_{ii}^2\right]\;,
\label{Helastic}
\end{equation}
where $\kappa$ is the vortex line curvature modulus and $\mu$ and
$\lambda$ are the Lam\'e coefficients.
$\mu$, $\lambda$ and $\kappa$ are
all expressible in terms of the parameters of the original model
Eq.\ref{F_GL}, and $u_{ij}={1\over2}(\partial_i u_j+\partial_j
u_i-\partial_z u_i\partial_z u_j)$ is the rotationally invariant {\em
nonlinear} strain tensor, with $i,j \in x,y$.  This form of $H_{el}$ is
strictly dictated by general symmetry considerations, specifically:
the underlying rotational symmetry of the SV solid about any arbitrary
axis 
guarantees that only the fully rotationally invariant
combination $u_{ij}$, given above, appears and that the vortex line tilt
modulus ($c_{44}$) vanishes identically (but see
below)\cite{unpublished}.

Although the softness of the SV lattice suggests, via the Lindemann
criterion, a lower melting temperature, our long wavelength
description cannot directly predict the melting temperature, because
thermal fluctuations are dominated by short distance modes (near the
Brillouin zone boundary).  Instead, here we focus on the much more
interesting, robust and universal effects of quenched disorder, all of
which (at long wavelengths) can be incorporated into our model by
adding to $H_{el}$
\begin{equation}
H_{dis}=\int d^3r\left[{\cal R}e \sum_{\bf G} V_{\bf G}({\bf r})e^{i {\bf G}
\cdot {\bf u}({\bf r})}+{\bf h}({\bf r})\cdot\partial_z{\bf
u}\right]\;.
\label{Hdisorder}
\end{equation}
The first term accounts for the random {\em positional} pinning (at
reciprocal lattice vectors ${\bf G}$) of the vortex lattice, with
$V_{\bf G}({\bf r})$ a complex random pinning potential, whose
statistics can be accurately represented by zero-mean, {\em
  short-ranged}, Gaussian correlations: $\overline{V_{\bf G}({\bf
    r})V_{\bf G'}^*({\bf r'})} = \Delta_V\delta_{{\bf G}{\bf
    G'}}^{\perp} \delta^d({\bf r}-{\bf r'})$\cite{RTaerogel}. The
second term reflects the presence of an additional {\em orientational,
  tilt} disorder.
The tilt field ${\bf h}({\bf r})$ is a random vector with {\em
short}-ranged {\em isotropic} spatial correlations, whose statistics
can also be taken zero-mean Gaussian, with $\overline{h_i({\bf r})
h_j({\bf r'})} =\Delta\delta_{ij}\delta^d({\bf r}-{\bf
r'})$\cite{RTaerogel}.  In the case of magnetic impurities, it is the
random local fields, acting on $\bf M$, i.e., ${\bf h}({\bf
r})\cdot{\hat{\bf n}}$, which via the Higg's mechanism, $\delta{\bf
n}\approx\partial_z{\bf u}$ lead to the tilt disorder piece of
$H_{dis}$.  In the case of nonmagnetic impurities, it is the
short-scale anisotropy in the spatial distribution of vortex line
pinning centers, together with lack of rotational symmetry in the SV
state, that leads to the tilt disorder in Eq.\ref{Hdisorder}, as can
be easily illustrated by short-scale modes thinning
procedure.\cite{unpublished} (As in conventional vortex lattices, in
$d>2$, the pinning of the {\em long-scale} vortex density distortions
is {\em irrelevant} at long scales).

Adapting Larkin's\cite{Larkin} calculation to the ``soft'' elasticity
of the SV lattice we find that in $d$ dimensions and at short length
scales, mean-squared lattice distortions (in domains of size
$L_\perp(\ll L_z)$) grow as $u_{rms}^2\sim L_\perp^{9/2-d}$. Hence,
translational order is destroyed for $d=3<d_c^{\Delta_V}=9/2$,
implying (in the weak disorder limit) divergently anisotropic Larkin
domains spanned by 3d Larkin lengths $L_\perp^L\approx\left(a^2(\kappa
  B_s^7)^{1/4}/\Delta_V\right)^{2/3}$ and $L_z^L\approx\left(a^2\kappa
  B_s/\Delta_V\right)^{1/3}$, with $a$ the vortex spacing, $s\in L,T$,
and $B_L=2\mu+\lambda$, $B_T=\mu$.  Beyond these Larkin length scales,
the Larkin approximation breaks down and the full nonlinear nature of
the random-field pinning potential must be taken into account, leading
to much more weakly (specifically, logarithmically) divergent lattice
fluctuations as in conventional vortex lattices\cite{FisherFRG}.  This
weak divergence is completely overwhelmed for $d<d_{uc}=7/2$ by the
far stronger fluctuations induced by the tilt disorder, which diverge
algebraically:
%
$u_{rms}^2\sim
L_\perp^{7/2-d}$ (ignoring nonlinear elastic effects) out to
arbitrarily long length scales
\cite{unpublished,RTaerogel,JSRTstretched,SRTcolumnar}. 
%

Ignoring nonlinear elastic terms, we obtain
the disorder-averaged phonon correlation function $\overline{C_s({\bf
r})}\approx \mbox{Max}[\Delta/(\kappa^3 B_s^5)^{1/4}r_\perp^{7/2-d},
\Delta/(\kappa^{5-d} B_s^{d-1})^{1/2}z^{7-2d}]\times O(1)$.
The spatial (Larkin-like) correlation lengths, measurable in neutron
scattering experiments, are defined by
$C_s(\mbox{Min}[\xi_\perp,\xi_z])=a^2$ and are given by $\xi_{\perp}
\approx\big[a^2(\kappa^3 B_s^5)^{1/4}/\Delta\big]^{1/(7/2-d)}$, $\xi_z
\approx\xi_\perp^{1/2}(\kappa/B_s)^{1/4}$. Hence, in 3d, in the 
presence of arbitrarily weak disorder $\Delta$, on length scales
longer than $\xi_{\perp,z}$ the translational order of SV solid decays
rapidly, that is, it is {\em short-ranged}.
%
%
This, however, does {\em not} imply that the system is equivalent to
the SV {\em liquid} state, or to a fully disordered vortex glass as it
would be if dislocation loops unbound. Using duality methods,
introduced recently by two of the authors\cite{RTaerogel}, we have
shown that a translationally disordered, but topologically ordered
``columnar elastic glass'' (CEG) phase is in fact possible in 3d.
This is analogous to our recent results in the context of randomly
pinned smectic liquid crystals\cite{RTaerogel,JSRTstretched} and
columnar phase of discotic liquid crystals\cite{SRTcolumnar} confined
in aerogel, as well as the Bragg glass conjecture for conventional
vortex lattices\cite{GL} and the random field XY model\cite{DSFisher}.

In this topologically ordered phase 
the anharmonic elastic terms in the Hamiltonian lead to drastically modified 
elastic scaling on length scales longer than the nonlinear crossover
length scales $\xi_{NL}^z=8\pi^2\kappa^2/\Delta$ and
$\xi_{NL}^\perp=(\xi_{NL}^z)^2(B_s/\kappa)^{1/2}$. In renormaliztion
group (RG) language, on scales longer than $\xi_{NL}^{z,\perp}$, our
system flows away from a zero-disorder Gaussian fixed point and
approaches a nontrivial $T=0$ anharmonic, disordered fixed point.
This new fixed point can
be studied using standard RG analysis\cite{RTaerogel}, and we predict
that at scales longer than $\xi_{NL}^{z,\perp}$, out to an arbitrary
length scale (but see below), the elastic moduli $\kappa$, $\mu$,
$\lambda$, and the disorder variance $\Delta$ become singular
functions of wavevector $\bf k$\cite{Brandt}:
\begin{mathletters}
\begin{eqnarray}
\kappa({\bf k})&=&\kappa_0\left(k_z\xi_{NL}^z\right)^{-\eta_\kappa}
f_\kappa(k_\perp\xi_{NL}^\perp/(k_z\xi_{NL}^{z})^\zeta)
\;,\label{kappa}\\
\mu({\bf k})&=&\mu_0\left(k_z\xi_{NL}^z\right)^{\eta_{\mu}}
f_\mu(k_{\perp}\xi_{NL}^{\perp}/(k_z\xi_{NL}^{z})^\zeta)
\;,\label{mu}\\
\lambda({\bf k})&=&g_*\mu_0\left(k_z\xi_{NL}^z\right)^{\eta_{\mu}}
f_\mu(k_\perp\xi_{NL}^\perp/(k_z\xi_{NL}^{z})^\zeta)
\;,\label{lambda}\\
\Delta({\bf k})&=&\Delta_0\left(k_z\xi_{NL}^z\right)^{1-2\eta_\kappa}
f_\Delta(k_\perp\xi_{NL}^\perp/(k_z\xi_{NL}^{z})^\zeta)
\;,\label{Delta}
\end{eqnarray}
\label{etas}
\end{mathletters}
where $f_{\kappa/\mu/\Delta}(x)$ is independent of $x$ if $x \ll 1$
and goes like $x$ to the power $-\eta_{\kappa}/\zeta$,
$\eta_{\mu}/\zeta$ and $(1-2\eta_{\kappa})/\zeta$, respectively, when
$x \gg 1$, so that, e.g., $\kappa({\bf k}) \propto
k_\perp^{-\eta_{\kappa}/\zeta}$ for $k_\perp\xi_{NL}^\perp \gg
(k_z\xi_{NL}^{z})^\zeta$.  $\kappa_0$, $\mu_0$ and $\Delta_0$ are the
bare values of the elastic constants and the anomalous exponents are
universal, and to leading order in $\epsilon={7\over2}-d$ are given by
$\eta_{\kappa}=1.478
\epsilon=0.74$ and $\eta_{\mu}=0.6919 \epsilon=0.346$, where the second
equalities are their values in 3d. The constant $g_*$ is also
universal, and to leading order in $\epsilon$ $g_*=-0.03272 +
O(\epsilon)$.  The anisotropy exponent $\zeta =
2-(\eta_{\kappa}+\eta_{\mu})/2$, and therefore the two independent
exponents $\eta_\kappa$ and $\eta_\mu$ completely characterize the
anomalous elasticity, with $\kappa({\bf k})$ and $\Delta({\bf k})$
diverging with linearly-related exponents, and $\mu({\bf k})$ and
$\lambda({\bf k})$ both vanishing in exactly the same way as ${\bf
k}\rightarrow\infty$. The negative universal amplitude ratio $g_*$
implies that the SV glass exhibits a {\em negative universal} Poisson
ratio $\sigma_p= -0.0338 + O(\epsilon)$.

For length scales longer than $\xi_{NL}$ this strong,
power-law anomalous elasticity will alter the behavior of
all physical observables of the SV solid, such as, e.g., the width of
the structure function peak $S({\bf k})$ measured in neutron
scattering and the behavior in its tails. It also leads to a
{\em non-Hookean}, i.e., nonlinear, stress-strain relation 
even for {\em arbitrarily small} applied stress $\sigma$.

To see this, consider a purely compressive stress,
$\sigma_{ij}=\sigma\delta_{ij}$,
which adds
a term $\sigma{\bbox\nabla}\cdot{\bf u}=\sigma u_{ii}+{1 \over
  2}|\partial_z{\bf u}|^2$ to the Hamiltonian.
In Fourier space, the second, symmetry breaking, term becomes $\sigma
k_z^2 |{\bf u}({\bf k})|^2$ and begins to dominate over the
$\kappa({\bf k}) k_z^4|{\bf u}({\bf k})|^2$ term once $\sigma k_z^2
\ge \kappa({\bf k})k_z^4$. This clearly happens for $k_z$'s less
than a critical $k_c$ given by $\sigma k_c^2=\kappa(k_z=k_c,{\bf
k}_{\perp}=0)k_c^4$. Using Eq.\ref{kappa} for $\kappa({\bf k})$ in
this expression and solving for $k_c$, we find
$k_c(\sigma)=(\sigma/\kappa_0)^{1/(2-\eta_{\kappa})}
(\xi_{NL}^z)^{\eta_{\kappa}/(2-\eta_{\kappa})}$.

Now, for sufficiently weak stress $\sigma$, $1/\xi^z_{NL}\gg
k_c(\sigma)$, the stress-induced rotational symmetry breaking term is
subdominant to the vortex curvature energy $\kappa({\bf k}) k_z^4|{\bf
u}({\bf k})|^2$ for wavevectors $k_z\gg k_c(\sigma)$. Consequently,
on intermediate length scales, $k_{c}(\sigma)<k_z<k_{NL}^z$, the SV glass
exhibits anomalous elasticity summarized by Eqs.\ref{etas}. However,
on longer length scales, $k_z\ll k_c(\sigma)$, the applied stress
energy dominates, suppressing fluctuations and cutting off
the anomalous elasticity at $k_c(\sigma)$. Therefore, the elastic
moduli {\em saturate}, for $k_z
\ll k_c(\sigma)$, at their values at $k_z=k_c(\sigma)$, $k_{\perp}=0$ given by
Eq.\ref{etas}, i.e., $\mu({\bf k}\rightarrow
0)\rightarrow\mu(k_z=k_c,{\bf k}_{\perp}= 0)\propto k_c^{\eta_{\mu}}$
and $\lambda({\bf k}\rightarrow 0)\rightarrow\lambda(k_z=k_c,{\bf
k}_{\perp}= 0)\propto k_c^{\eta_{\mu}}$. This implies that the Young's
modulus $Y(\sigma) = 4\mu(\mu +\lambda)/(2\mu +\lambda)=
4\mu_0(1+g_*)/(2+g_*)(k_c\xi_{NL}^z)^{\eta_\mu}\approx
2\mu_0(k_c\xi_{NL}^z)^{\eta_\mu}=2\mu_0(\sigma/\sigma_{NL})^\beta$,
with $\sigma\ll\sigma_{NL}$, $\beta=\eta_\mu/(2-\eta_\kappa)$, and
$\sigma_{NL}=\kappa_0\xi_{NL}^{-2}$.  Hence, we find for stress
$\sigma\le\sigma_{NL}$, a {\em nonlinear} (non-Hookean) strain-stress
relation: $\epsilon_c(\sigma)\equiv\bbox{\nabla}\cdot{\bf
u}=\sigma/Y(\sigma)\propto
\sigma^{\alpha}$, with $\alpha=1-\beta=1-\eta_{\mu}/(2-\eta_{\kappa})
\approx 0.72$.

Now, it is easy to see that ${\bf H}=H\hat{z}$, applied {\em along}
the direction of the SV lattice, acts as a compressive stress,
$\sigma=H B(0)$, and the induced $\delta B(H)=B(H)-B(0)$ plays the
role of the compressive strain, $\epsilon_c=\delta B/B(0)$. Hence $H$
allows us to {\em directly} probe the anomalous elasticity,
with the non-Hookean elasticity leading to the nonlinear and universal
$B(H)$ relation, for
$H<H_{NL}=\sigma_{NL}/B(0)$, given in Eq.\ref{nonhook} and Fig.\ref{BvsH}.


Of course, as with all crystalline ferromagnets, crystalline
symmetry breaking fields will {\em explicitly} break rotational
invariance, even in the absence of ${\bf H}$. They will lead to a
non-zero tilt modulus, i.e., a term ${1\over2}V_{cr.}|\partial_z{\bf
u}|^2$ in the Hamiltonian, which will cut off the anomalous elasticity
on length scales longer than $\xi_{cr.}^z=({\kappa_0 \over
V_{cr.}})^{1/(2-\eta_\kappa)} (\xi_{NL}^z
)^{-\eta_{\kappa}/(2-\eta_\kappa)}$. For weak $V_{cr.}$, our
predictions will apply over a wide range of length scales
$L_{z,\perp}$ satisfying $\xi_{cr.}^{z,\perp} \gg L_{z,\perp} \gg
\xi_{NL}^{z,\perp}$.  On longer length scales the SV solid elasticity
will cross over to the conventional ``tension'' elasticity of vortex
lattices\cite{ReviewsSC}, exhibiting a {\em linear} relation $\delta B
\propto H$, for $H<H_{cr.}=V_{cr.}/B(0)$.

The crystal anisotropy also leads to metastability for
$|H|< H_{cr.}$, thereby allowing experimental studies of $\bf H$
applied in the direction {\em opposite} to that of $\bf M$. In the
limit $|H|\rightarrow H_{cr.}$, the vortex solid becomes ``soft''
again, with the crystal-field generated tilt modulus vanishing as
$H_{cr.}-|H|$, and the system exhibiting anomalous elasticity out to
arbitrary length scale at this finely field-tuned point. However,
because the rotational symmetry is nevertheless broken (with
restoration only at the quadratic $(\partial_z {\bf u})^2$ level), we expect
that the universality class and therefore the anomalous exponents will be
distinct from that of the CEG studied here.\cite{unpublished}

As argued in the case of vanishing crystal anisotropy, here too, the
anomalous elasticity leads to a non-trivial field dependence of the
Young's modulus.  We find $Y(H)\propto(H_{cr.}-|H|)^{\beta'}$, with
$\beta'>0$ a universal exponent analogous to, but distinct from the
$\beta$ of the CEG. However, in contrast to the rotationally invariant
case, here the flux density $\delta
B(H)=H/Y(H)=H(H_{cr}-|H|)^{-\beta'}$ diverges as $|H|$ approaches
$H_{cr.}$ from below. This signals a breakdown of the analysis, which
assumes $\delta B(H)\ll B(0)$, and suggests that the metastability
limit occurs at fields $H_{ms}$ determined by $\delta B(H_{ms})\approx
B(0)$.

Finally, we note that ErNi$_2$B$_2$C has a strong uniaxial anisotropy,
with $\bf M$ preferentially lying in a nearly isotropic xz-plane. The
appropriate model in this case resembles that given by
Eq.\ref{Helastic}, except that in the strain tensor $u_{ij}$,
$u_i\rightarrow u_x$, and $u_y$ has a conventional quadratic, tension
elasticity. A similar model has recently been considered by us in the
context of the columnar phase of discotics confined in an anisotropic
random medium, such as, e.g., strained aerogel\cite{SRTcolumnar}. The
resulting ``$m=1$ elastic glass'' phase is also topologically stable
with anomalous elasticity, and has additional interesting features,
most notably, that it exhibits short-ranged translational order in one
direction ($x$) and {\em quasi}-long-ranged translational order in the
other ($y$), a prediction that should be testable via neutron
scattering and decoration experiments. We conclude by noting that the
``soft'' elasticity of the SV solid should also manifest itself in
many other physical observables, most notably in dynamics, exhibiting
novel current-voltage characteristics and enhanced critical currents,
sensitive to and with nontrivial dependence on $H$.

We thank Anton Andreev and Tom Lubensky for discussions and
acknowledge financial support by the NSF through DMR99-80123 (JT,KS),
DMR96-25111, MRSEC DMR-9809555, and the A.P. Sloan and The
David and Lucile Packard Foundations (LR,ME).
%

%
\end{multicols} 
\end{document}